\documentclass[10pt,conference]{IEEEtran}

\usepackage[table]{xcolor} 
\usepackage{amsmath}
\usepackage{subfig}
\usepackage{pdfpages}
\usepackage{tikz}
\usepackage{svg}
\usepackage{soul}
\usepackage{cleveref}
\usepackage[multiple]{footmisc}
\usepackage{rotating}
\usepackage{textcomp}
\usepackage{amssymb}
\usepackage{xurl}

\usepackage{glossaries}

\newacronym{iot}{IoT}{Internet of Things}
\newacronym{pcb}{PCB}{Printed Circuit Board}

\AtBeginDocument{%
  \providecommand\BibTeX{{%
    \normalfont B\kern-0.5em{\scshape i\kern-0.25em b}\kern-0.8em\TeX}}}

\begin{document}

\title{Observing a Moving Target - Reliable Transmission of Debug Logs from Mobile Embedded Devices}

\author{\IEEEauthorblockN{Björn Daase, Leon Matthes, Lukas Pirl, Lukas Wenzel}
\IEEEauthorblockA{Hasso Plattner Institute, University of Potsdam, Germany\\
\{bjoern.daase, leon.matthes\}@student.hpi.de, \{lukas.pirl, lukas.wenzel\}@hpi.de}}

\newcommand{\todobase}[3][RGB]{
  \begingroup%
  \definecolor{hlcolor}{RGB}{#2}\sethlcolor{hlcolor}%
  \hl{#3}%
  \endgroup}

\newcommand{\todo}[1]{\todobase{255, 255, 0}{#1}}
\newcommand{\bd}[1]{\todobase{124,252,0}{bd: #1}}
\newcommand{\jb}[1]{\todobase{124,252,0}{lm: #1}}
\newcommand{\lp}[1]{\todobase{255,0,255}{lp: #1}}

\renewcommand\figurename{Figure}
\renewcommand\tablename{Table}
\renewcommand\thetable{\arabic{table}}
\newcommand{\textapprox}{\raisebox{0.5ex}{\texttildelow}}

\rowcolors{3}{lightgray!30}{white}

\maketitle

\begin{tikzpicture}[remember picture,overlay]
  \node[anchor=south,yshift=10pt] at (current page.south) {\fbox{\parbox{\dimexpr\textwidth-\fboxsep-\fboxrule\relax}{
    \footnotesize \textcopyright 2021 IEEE. Personal use of this
    material is permitted. Permission from IEEE must be obtained for
    all other uses, in any current or future media, including
    reprinting/republishing this material for advertising or
    promotional purposes, creating new collective works, for resale or
    redistribution to servers or lists, or reuse of any copyrighted
    component of this work in other works.
  }}};
\end{tikzpicture}%

\begin{abstract}

Mobile embedded devices in the \acrlong{iot} face tight resource constraints and uncertain environments, including energy scarcity and unstable connectivity.
This aggravates debugging, optimization, monitoring, etc.; for which logs from devices are required throughout the whole product life cycle.

In this work, we qualitatively compare approaches for transmitting logs with regard to application requirements (e.g., continuous or aperiodic transmission), resource consumption (e.g., memory), operating constraints (e.g., power supply), and the transmission medium (e.g., UART, WiFi).
The comparison highlights that the appropriateness of the approaches further depends on the life cycle phase and the implementation effort of an approach.

We report on a case study in which we developed a mobile embedded device (i.e. a self-driving slot car) with which logs and firmware have to be exchanged.
Compared to wired transmission, WiFi has shown to be more flexible and suitable for more phases in the development process.
However, additionally required computations and energy distort the behavior of the resource-constrained device.
E.g., when transmitting logs wirelessly, the device's supply voltage drop within a \textapprox20~ms power interruption is \textapprox1.6 times higher than without a loaded WiFi stack.
Thus, also debugging must be considered in the energy budgeting.

\end{abstract}
\section{Introduction}
With over 98\% of all manufactured microcontrollers being embedded microcontrollers~\cite{ebert2009embedded}, developing and operating such devices is a major area of responsibility.
Around 50\% of software development time is spent for making code work and resolving defects (commonly referred to as \emph{bugs}) \cite{britton2013reversible}. For the sake of readability, this work refers to the aforementioned activities, identifying performance and hardware issues, etc. commonly as \emph{debugging}.

Debugging is not limited to the development process but might also be required in further phases of a device life cycle.
The phases of a device life cycle encompass different debugging scenarios with different characteristics and constraints.
For example, the development process on the one hand and monitoring in production on the other hand have completely different challenges (i.e. easily adaptable debugging vs. constant debugging output).

The deeper a device is embedded, the harder the debugging.
While common debugging tools require physical access to the device to transfer debug logs, deeply embedded devices are normally not physically accessible~\cite{DBLP:journals/sigbed/Mohan08}.
The situation is even more difficult when the embedded device is also mobile.

In this work, we focus on exploring different debugging techniques for mobile embedded devices.
We also analyze how these techniques map to different phases of a device's life cycle and how they can improve the debugging experience.
To assess the applicability of the different techniques in practice, we conduct a case study with self-driving slot cars.
In addition to the analysis and case study, we confirm previous research~\cite{DBLP:conf/isorc/RichterGP15} in that our lab setting is suited for imitating real-world challenges regarding the development and operation of embedded devices.
\section{Background and Related Work}
\label{sec:Background_and_Related_work}

In this section, we first give an overview of the challenges when debugging mobile embedded devices in \Cref{subsec:Debugging_Mobile_Embedded_Devices}.
\Cref{subsec:Debugging_Scenarios} describes the various scenarios that are derived from the different facets of debugging.
Constraints imposed by mobile embedded devices are outlined in \Cref{subsec:Constraints}.

\subsection{Debugging Mobile Embedded Devices}
\label{subsec:Debugging_Mobile_Embedded_Devices}

Debugging embedded systems is known to be a challenging task~\cite{macnamee2000emerging}, often also adding up constraints from, e.g., (hard) real-time and distributed systems~\cite{DBLP:journals/csur/Stankovic96}.
In the past, researchers have analyzed the debugging of embedded systems in various contexts, e.g., for wireless embedded networks~\cite{DBLP:conf/ipsn/WhitehouseTTSKJHDC06}, multimedia applications~\cite{DBLP:conf/emsoft/CuevaBTMS12}, or on ships~\cite{MacNamee2000EmergingOD}.
As these domains highly differ, the understanding of what debugging includes is different~\cite{IBM:journals/Hailpern02} as well.
This suggests grouping debugging scenarios into different categories with specific requirements.

Data transmission is the key to debugging, as only through communication device states can be determined and diagnosed~\cite{murali2021improved}.
The communication technologies commonly used for embedded devices tend to be different and more diverse compared to those of larger computer systems~\cite{DBLP:conf/codes/SamieBH16}.
Aspects such as resource consumption and timing behavior tend to be of a stronger concern than, e.g., throughput.
At the same time, the wide variety of communication technologies used in embedded systems can be considered characteristic.
In the past, both wired and wireless transmission technologies have been used for transmitting data from embedded devices~\cite{DBLP:conf/codes/SamieBH16, yue2013marine}.
Regarding wireless transmission, there is a large body of research on Bluetooth and WiFi~\cite{yue2013marine, DBLP:conf/infocom/FriedmanKK11}.
Especially for WiFi, extensive experiments on energy consumption have already been carried out~\cite{DBLP:conf/i2mtc/PotschBS17}.

Mobile embedded devices are increasingly deployed in mission-critical applications~\cite{DBLP:conf/hpec/SkowyraBB13}.
The mobility of these devices yields two major challenges:
First, mobile embedded devices might not have a stable energy supply.
Therefore, the energy usage might impact the stability of the system.
At least transiently, energy capacities might be reduced and might even be too little to keep a system functional~\cite{DBLP:conf/aspdac/HwangKJC08}.
Second, not every transmission technology is suitable for every environment. 
While there is a lot of research focusing on the security of mobile embedded devices~\cite{ DBLP:conf/hpec/SkowyraBB13, 5751382, 10.1145/1435458.1435462}, choosing an adequate transmission technology is also dependent on the environment, specifically on energy requirements.

\subsection{Debugging Scenarios}
\label{subsec:Debugging_Scenarios}

Previous research on debugging embedded systems has focused in particular on the development cycle performed by hardware and software developers~\cite{DBLP:conf/uist/DrewNMMMH16, DBLP:conf/chi/BoothSBJ16, DBLP:conf/uist/McGrathWKHDH18, DBLP:conf/osdi/FonsecaDLS08}.
However, as Hailpern et al. point out~\cite{IBM:journals/Hailpern02}, debugging is not limited to software development, but also extends into other parts of the development cycle, including production.
All these different activities require their own capabilities to allow for effective debugging.
In this section, we highlight core development and operation activities during which debugging is required and characterize their requirements respectively.

\subsubsection*{Software Development}
During development, debugging is essential to ensure that the written code does not contain any faults~\cite{IBM:journals/Hailpern02}.
Developers need to conduct experiments to assess how a target platform reacts to certain commands and operating states.
During development, code changes frequently, required debugging output changes frequently, and every hour of development is expensive.
Hence, debugging facilities must be available straightforward, easily adaptable, and must allow for quick code iteration~\cite{DBLP:journals/isj/BaskervilleP04}.

\subsubsection*{Optimization}
Optimization activities increase in the later stages of development.
This is an especially important phase in embedded systems development, where hardware and software tend to be closely coupled.
The hardware is typically close to production-ready, so optimization efforts can take hardware-specifics into account.
The Software is typically already functional but needs to fulfill tighter resource constraints.
As a result, the analysis focuses on identifying capacity and performance bottlenecks as code changes become less frequent~\cite{DBLP:journals/ubiquity/Hyde06}.
Therefore, the debugging instrumentation should have a small and predictable runtime impact.
Ideally, the debugging and optimization techniques are useful for (hard) real-time systems as those systems require minimal impact on program behavior.

\subsubsection*{Monitoring}
During and after development, gaining insight into the embedded systems' states is important to identify misbehaving products and unexpected environmental influences.
For this reason, constantly monitoring certain aspects of embedded devices might be required~\cite{DBLP:journals/spe/DoddR92}.
This can include environmental sensors, hardware states, or dynamic software analyses.
During development, this data should be available quickly, ideally in a continuous stream.
In production, data could be limited to aggregated monitoring reports which are transmitted periodically or logged locally for access in the event of unexpected behavior of the device.
Continuous output is usually not necessary in these scenarios.

\subsubsection*{Operational Status}
The operational status can be derived from monitoring data and summarizes the condition of a device (e.g., commands executed, data processed, resource usage)~\cite{DBLP:journals/cacm/Spinellis18}.
Whilst throughput is of lesser concern, such statuses need to be reported with low latency, as they can indicate problems with the device or critical operating conditions.

\subsubsection*{Defects in Production}
Despite the best efforts of developers, defects in the code might reach the production phase.
As it is unavoidable for humans to err, responsible developers must take measures to be able to detect and debug defects in production code~\cite{IBM:journals/Hailpern02}.
At this point, the distance (physical, as well as virtual) to the device can be challenging.
In case the developer is unable to access a device physically, the transmission of data required for debugging must be automated or made easy for a user to do so.
Additionally, means to deploy new software versions need to be in place.

\subsection{Debugging Constraints}
\label{subsec:Constraints}

Embedded devices are characterized by low computational power as well as little memory availability.
In the case of mobile embedded devices, energy consumption must also be considered. This is of concern especially for battery-operated devices, where increased computational efforts entail increased energy consumption.
Moreover, mobile embedded devices might not always be physically accessible, which can limit the duration, time, and locality for debug data transfer.
Furthermore, if such a device is used in a (hard) real-time system, predictability is also required.

\section{Debugging Techniques Comparison}
\label{sec:Debugging_Techniques_Comparison}

In \Cref{subsec:Debugging_Mobile_Embedded_Devices}, we have described that embedded devices can be differentiated by the used communication medium.
The use of either a wired or wireless transmission media can be seen as a major point of differentiation.
For mobile embedded devices, the availability of a communication medium and the characteristics of the energy supply, suggest an additional differentiation between a continuous or aperiodic data transmission. 
In this section, we therefore derive a matrix of debugging techniques from the constraints given in \Cref{subsec:Constraints}.

\begin{figure}
    \vspace{-0.3cm}
    \centering
    \includegraphics[width=0.75\linewidth]{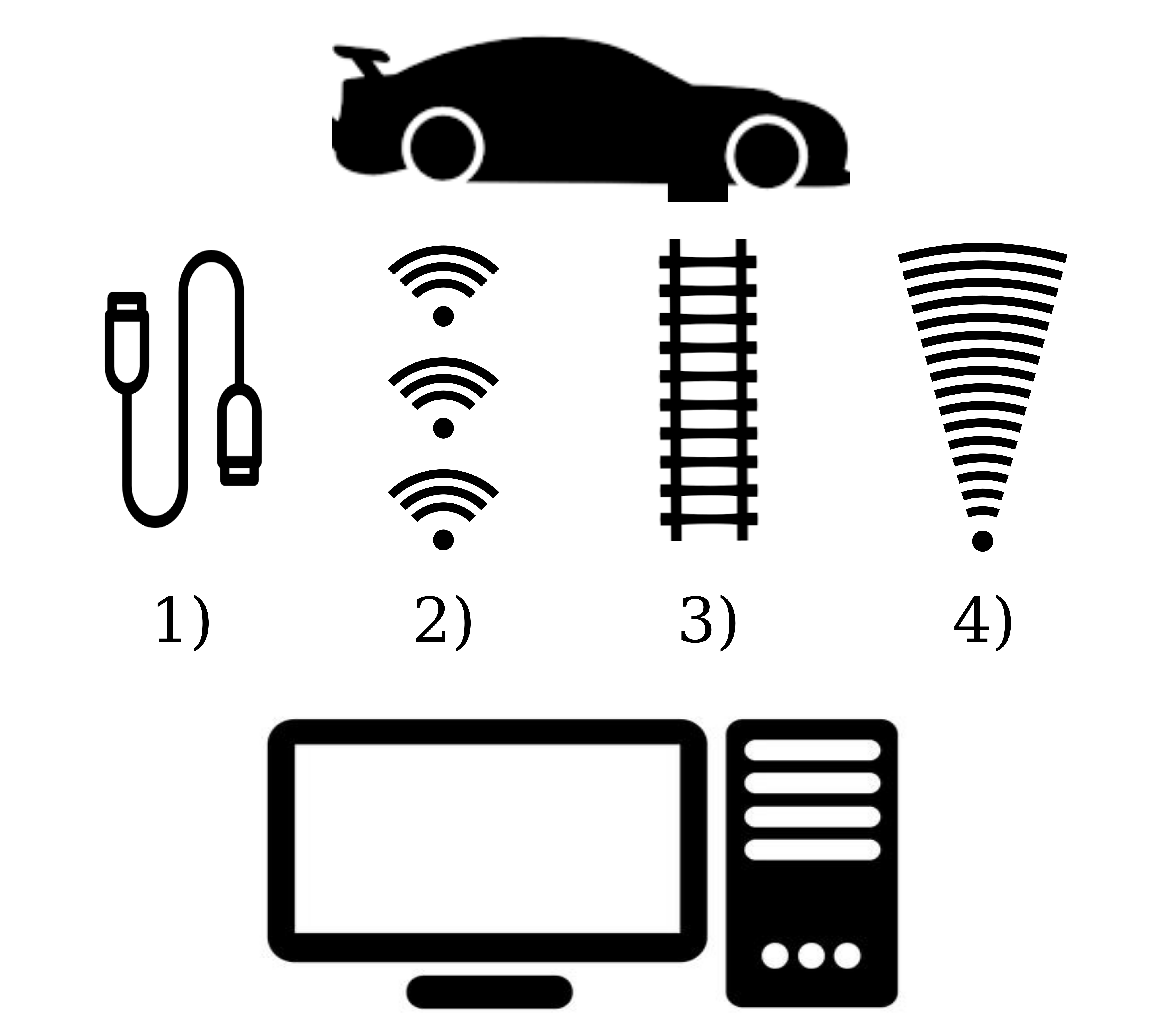}
    \caption{Debugging Techniques: 1) Aperiodic Wired Transmission 2) Aperiodic Wireless Transmission 3) Continuous Wired Transmission 4) Continuous Wireless Transmission.}
    \vspace{-0.2cm}
  \label{figure:Debugging_Techniques}
\end{figure}

We subdivide the available debugging techniques for embedded devices into four categories along the axes of communication media, as well as their availability as shown in \Cref{figure:Debugging_Techniques}.
For each of the four categories, debugging techniques are introduced and assessed qualitatively.
A case study evaluating these four techniques on the example of self-driving slot cars will be presented in the subsequent \Cref{sec:Case_Study}.

\subsubsection{Aperiodic Wired Transmission}
In this case, stable wired transmission is available only during limited periods of time or in certain operating conditions.
It requires saving debugging logs until the wired connection is available.
This results in an increased memory and storage consumption.
Additionally, increased computational resources are required whenever the connection becomes available.

The low computational overhead during logging makes it especially applicable for optimization purposes.
The typically low setup and implementation efforts needed for implementing a wired transmission make it applicable for early stages of development as well.

An example could be a solar-powered environmental sensor where the energy supply is insufficient for wireless transmission but where data is being regularly collected on-site.

\subsubsection{Aperiodic Wireless Transmission}
When wireless transmission media are available, however not continuously and stable, then the use of an aperiodic wireless transmission can be applicable.
Again, this requires saving debugging logs until wireless communication is established.
Therefore, more memory and storage are required, as well as additional computational overhead when establishing a wireless connection and transmitting.
Especially the setup and teardown of the wireless communication stack (e.g., WiFi/Bluetooth) can incur a significant overhead~\cite{DBLP:conf/www/QianWGHGMSS12}.

Especially in a development environment, an aperiodic wireless transmission can be stably ensured which makes this technique usable for remote software development.
Likewise, due to the low overhead when not transmitting, this technique can be suitable for optimization as well.
The ease of use and widespread adoption of wireless technologies makes this technique usable also in production for transmitting failure reports alike.

In the real world, wearable devices (e.g., health trackers, smartwatches) fit in this category, as they often require proximity to the home network or a smartphone for Internet access.

\subsubsection{Continuous Wired Transmission}
For this technique, continuous access to a wired connection must be ensured.
In comparison to wireless equivalents, lower overhead in computation and data storage can be expected.
Because the physical setup of a separate infrastructure solely for the transfer of debugging logs is expensive and often infeasible, existing infrastructures can be reused (e.g., power lines).

The overall low overhead makes this technique satisfactory for monitoring and operational status.
These require continuous but small data transfers.
Additionally, the low overhead makes it usable for optimization purposes as long as the limited data rates are sufficient.

Whilst this case is theoretically possible, it is also the most unlikely to occur in practice, as it would require continuous access to a mobile device via a wired transmission medium.
However, one such case could be the transmission of data using the contact wire of a railway track.

\subsubsection{Continuous Wireless Transmission}
Wireless transmission has the advantage of easier availability for devices on the move.
On the flip side, the continuous operation of the wireless stack is compute- and thus energy-intensive.
However, data does not have to be stored for extended periods on the device.

The continuous connection is especially advantageous when transmitting monitoring and operational status data.
If a stable wireless transmission can be ensured, this technique is applicable for development purposes.
Due to its ease of access, it encourages faster development cycle times that improve the quality of development~\cite{DBLP:journals/isj/BaskervilleP04}.
Furthermore, a failure in production can more easily be reported, as wireless networks are more and more part of production environments.

This makes it usable, e.g., for smartphones that provide continuous access to wireless (including cellular) networks.
\section{Case Study - Self-Driving Slot Car Debugging}
\label{sec:Case_Study}

Previous research has shown that there can be considerable gaps between theoretical and experimental results for transmission communication media~\cite{DBLP:conf/etfa/SenoVT11}.
Especially the power consumption of our custom \gls{pcb} in interplay with the environment-dependent power consumption of wireless protocols is expected to be difficult to simulate representatively.
Hence, we abstained from simulations and conducted a hard- and software implementation directly.

This section first describes our experiment setup using slot cars in \Cref{subsec:Experiment_setup} and explicitly highlights the energy consumption constraints of our experiment in \Cref{subsec:Energy_Consumption}.
Then, we apply the techniques described in \Cref{sec:Debugging_Techniques_Comparison} to our slot car in \Cref{subsec:Debugging_Technique_Application}.
Finally, \Cref{subsec:Evaluation} evaluates the different approaches.

\subsection{Experiment Setup}
\label{subsec:Experiment_setup}

\begin{figure}
    \vspace{-0.1cm}
    \centering
    \includegraphics[width=6cm]{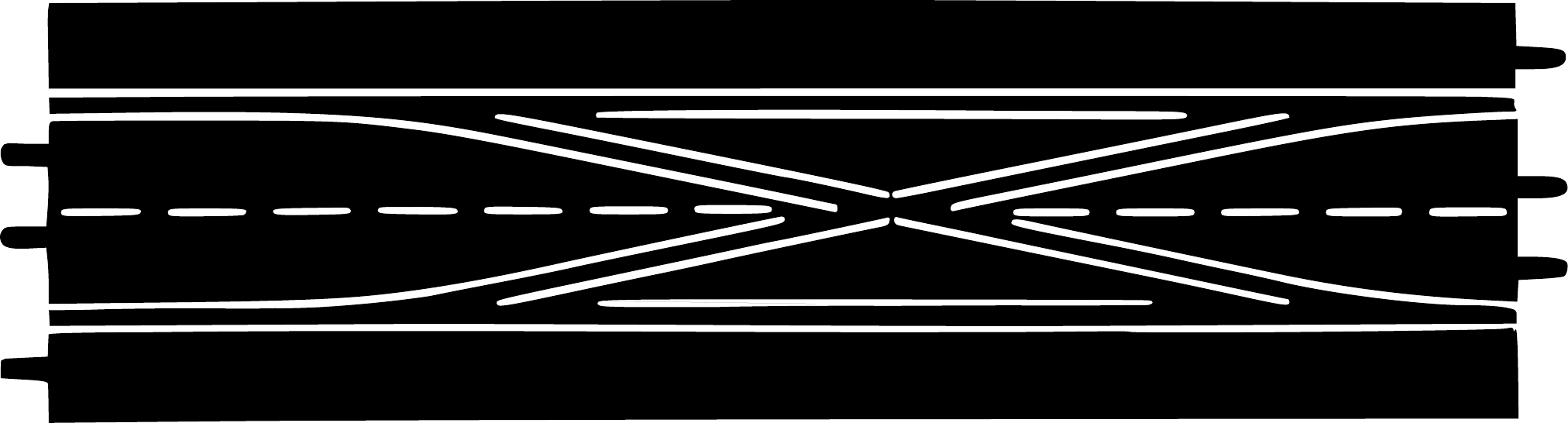}
    \caption{Double lane change track segment.}
    \vspace{-0.3cm}
  \label{figure:double_lane_chnage}
\end{figure}

In a setup similar to previous research~\cite{DBLP:conf/isorc/RichterGP15}, we use a digital Carrera D132 race track which resembles an environment with typical aspects for the development of mobile embedded devices.
To test the techniques described in \Cref{sec:Debugging_Techniques_Comparison}, we develop a custom slot car controller based on the ESP32 microcontroller.
This allows us to run custom code on a moving slot car and thereby imitate a remote, mobile, and difficult-to-reach embedded device.

Initially, we connect multiple development boards on a board for hardware prototyping (i.e., \emph{perfboard}).
This setup is easy to program and communicate with, using the built-in USB port of the ESP development board.
It is a typical prototyping scenario for embedded devices.
The later revision is implemented as a custom \gls{pcb} with all components integrated and UART access for programming and debugging.
The hardware shows that even relatively simple problems require relatively complex hardware solutions.
Due to the complexity, purely theoretical considerations, such as simulations, cannot easily cover the functionality in interaction with environmental influences.
We assume that this board resembles a product close to market, where development and debugging access is more difficult.
Especially because this \gls{pcb} fits into the chassis of a Carrera slot car, reaching the UART port requires disassembling the car even in the lab.

The slot car is continuously supplied with power over two metal rails (i.e., also when not moving) except on lane change segments, where the rails are interrupted as shown in \Cref{figure:double_lane_chnage}.
On these segments, the slot cars must drive (i.e., roll) up to \textapprox6~cm without a power supply.
These power interruptions must be considered when programming the microcontroller, as program execution has a relevant impact on power consumption.

The overall aim of the development is a self-driving slot car.
It is important to note that the slot car is not supposed to represent an autonomous vehicle.
The debugging data in our experiment form logging data and firmware, which need to be exchanged between the development host and the slot car during the development.
During the power interruptions on lane change segments, energy is limited to the capacity of a 1000~\textmu F capacitor.
The car can sustain normal operation of the microcontroller (clock speed 80~MHz, no WiFi) for up to 100~ms.
Motor power is cut off during these segments, as the motor control is separate from the ESP's power circuitry.

\subsection{Energy Consumption}
\label{subsec:Energy_Consumption}

\begin{figure*}%
    \vspace{-0.3cm}
    \centering
    \subfloat[\centering 80~MHz clock frequency, No WiFi.]{{\includegraphics[width=5cm]{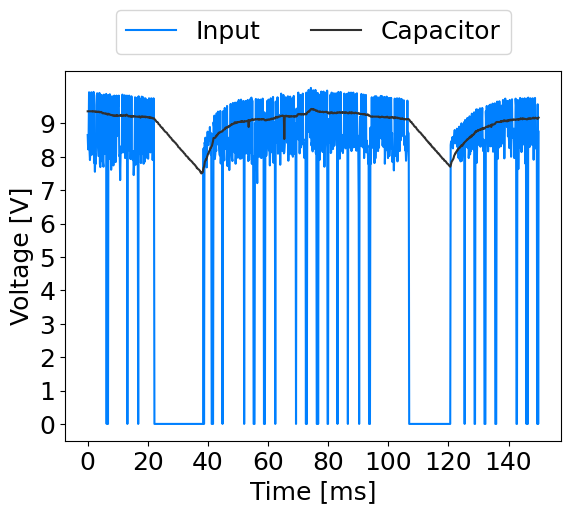} }}%
    \qquad
    \subfloat[\centering 80~MHz clock frequency, Not Sending.]{{\includegraphics[width=5cm]{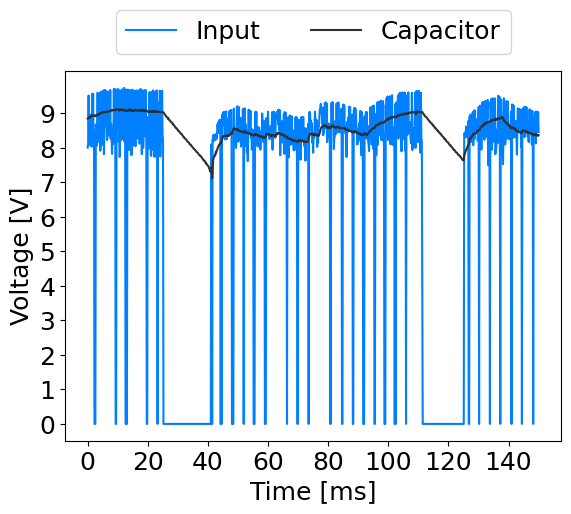} }}%
    \qquad
    \subfloat[\centering 80~MHz clock frequency, Sending.]{{\includegraphics[width=5cm]{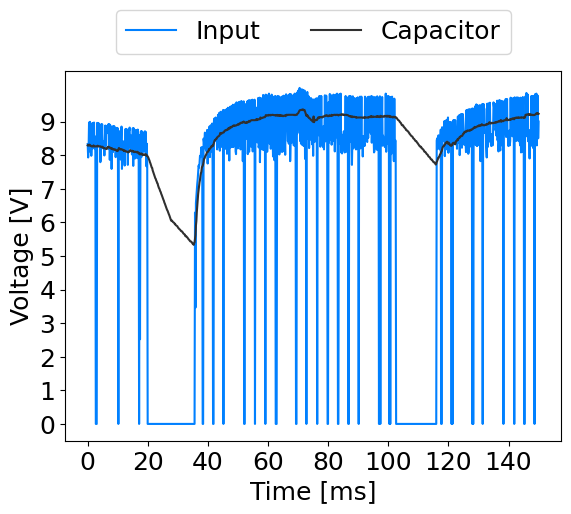} }}%
    \qquad
    \subfloat[\centering 160~MHz clock frequency, No WiFi.]{{\includegraphics[width=5cm]{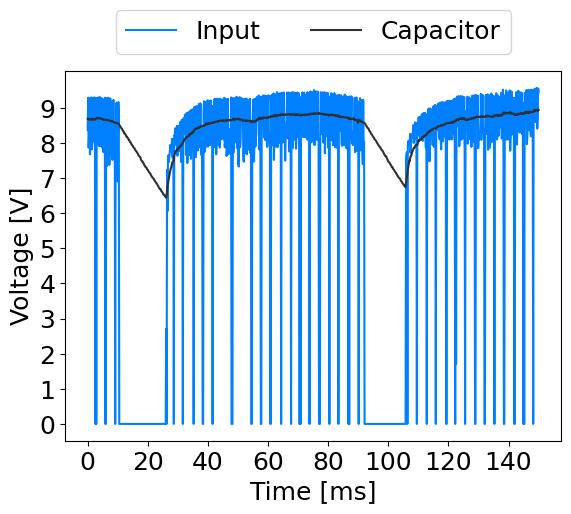} }}%
    \qquad
    \subfloat[\centering 160~MHz clock frequency, Not Sending.]{{\includegraphics[width=5cm]{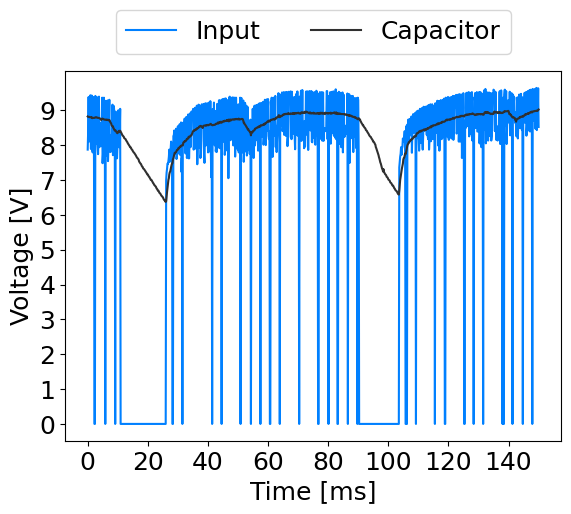} }}%
    \qquad
    \subfloat[\centering 160~MHz clock frequency, Sending.]{{\includegraphics[width=5cm]{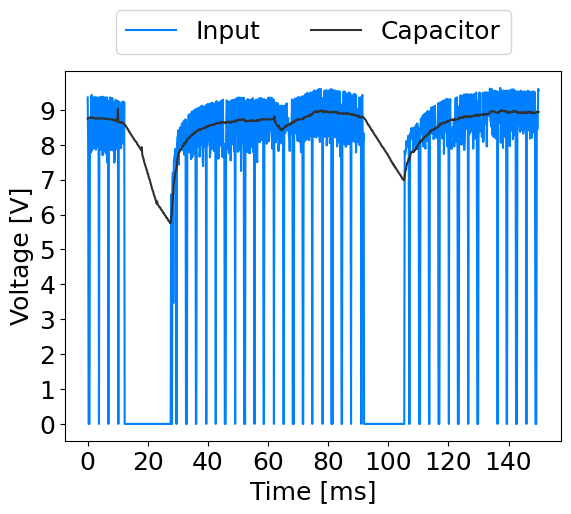} }}%
    \caption{Continuous voltage readout of the track (input) and of the capacitor.}%
    \label{figure:Power_Consumption}%
\end{figure*}

\begin{figure}
    \vspace{-0.3cm}
    \centering
    \includegraphics[width=5cm]{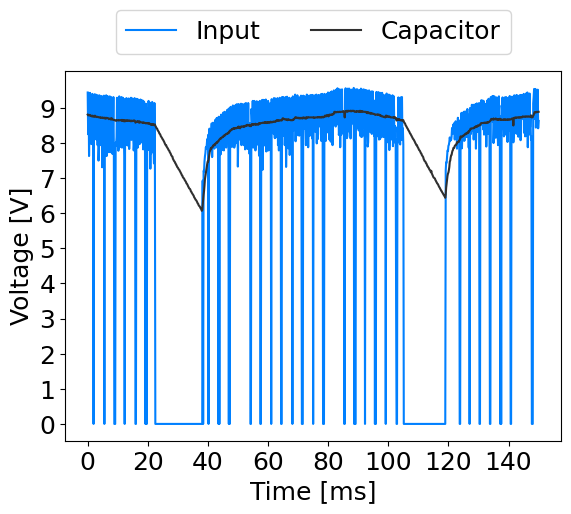}
    \caption{240~MHz clock frequency, No WiFi.}
    \vspace{-0.3cm}
  \label{figure:240MHz}
\end{figure}

The energy consumption of the ESP32 is influenced by multiple factors.
During the crossing of a lane change segment, it needs to be tightly controlled, as no power supply is available.
Apart from putting the microcontroller into a sleep mode that prohibits calculation, changing the operating frequency, and limiting the use of wireless communication has a significant impact on the consumed power.

\Cref{figure:Power_Consumption} illustrates the supply and capacitor voltages during the two power interruptions of a lane change segment (at \textapprox 30~ms and \textapprox 120~ms).
The first row, represents the microcontroller operating at 80~MHz clock frequency, the second at 160~MHz.
Individual columns represent (from left to right) operation without the WiFi stack active, with the WiFi stack active but not sending, as well as WiFi active and sending.
The microcontroller is forced to send in column three by repeatedly requesting data from it. 
Requests are timed in a way that they have to be answered exactly while the slot car is on a lane switch segment.

As the supply voltage also encodes the Carrera digital communication protocol, as described in \Cref{subsubsec:Write_to_Carrera_Track}, it fluctuates even during normal operation.
The maximum observed capacitor voltage drop in all scenarios is shown in \Cref{table:Voltage_Drop}.
Where the microcontroller consistently failed operation due to brownouts (capacitor voltage drop $\gtrapprox$~4~V), \textit{n/a} is listed.

Should the capacitor voltage drop below 5~V (the operating voltage of the ESP32) at any point, the brownout detector of the microcontroller detects that, resulting in an automatic reboot of the ESP32 and its peripherals.
As a consequence, data stored non-permanently would be lost and the self-driving would be impaired.
It is apparent that these voltage drops need to be tightly controlled to avoid reboots of the microcontroller.

As also shown in previous work~\cite{DBLP:journals/tvlsi/ZhaiPNHORMHASB09} and hence expected, the operating frequency influences the rate at which the voltage of the capacitor drops as seen in \Cref{figure:Power_Consumption} a) and d) and \Cref{figure:240MHz}.
Furthermore, \Cref{figure:Power_Consumption} b) and e) show that enabling WiFi without actually sending or receiving data slightly increases energy consumption already.
Apparently, the activated WiFi driver stack and chip result in an increased power consumption of 95\textapprox100~mA~\cite{ESP32DataSheet}.

Sending WiFi signals however, is the most problematic scenario as demonstrated in \Cref{figure:Power_Consumption} c) and f) as it draws \textapprox240~mA~\cite{ESP32DataSheet}.
The capacitor voltage drops rapidly in both instances for a very short period of time.
This happens when the microcontroller answers a request from the development host exactly during the first power interruption in both figures.
When using the microcontrollers' maximum frequency of 240~MHz, power consumption is increased again, as shown in \Cref{figure:240MHz}.
The use of WiFi at 240~MHz is infeasible as the car could not cross a lane change segment without brownouting.

\begin{table}
    \centering
        \begin{tabular}{c | ccc}
            & No WiFi & Not Sending & Sending\\
            \hline
            80~MHz & 1.62~V & 1.91~V & 2.64~V\\
            160~MHz & 2.11~V & 2.20~V & 2.82~V\\
            240~MHz & 2.49~V & \textit{n/a} & \textit{n/a}
        \end{tabular}
    \caption{Maximum observed capacitor voltage drop.}
    \vspace{-0.3cm}
    \label{table:Voltage_Drop}
\end{table}

\subsection{Debugging Technique Application}
\label{subsec:Debugging_Technique_Application}

This section describes the application of the debugging techniques described in \Cref{sec:Debugging_Techniques_Comparison}.
After an introductory description of each technique, we discuss their characteristics, give a recommendation for fitting debugging scenarios from \Cref{subsec:Debugging_Scenarios}, and finally evaluating them.

\subsubsection{Save and Print Later (Aperiodic Wired Transmission)}
While driving, the slot car continuously measures data and writes it to internal flash memory.
After a certain period of time, the car stops at a predefined point (e.g., at the start/finish line) and is connected to a read-out device (usually a laptop computer) via a cable.
Then, the car transmits its data via the cable.
At this point, there is also the option to reprogram the car.
When finished, the cable is disconnected, the car deletes the stored data and drives off again.

\textbf{Advantages}
\begin{itemize}
    \item Known method --- easy to set up
    \item Low runtime performance impact
    \item Very low risk of interruption/data corruption
\end{itemize} 

\textbf{Shortcomings}
\begin{itemize}
    \item Stopping not possible everywhere, some parts of track not accessible
    \item Needs physical access, which is difficult inside chassis and UART header is difficult to connect to
    \item Stopping interrupts normal operation --- possibly infeasible in production
    \item Data must be saved whilst driving
\end{itemize}

\textbf{Evaluation}
Overall, "Save and Print Later" is a very simple and reliable form of transferring data from the car as well as uploading firmware to the car.
On the other hand, physical access to the board in the car is required.
The deeper a device is embedded, the more difficult it becomes, and size restrictions also limit the use of the transmission technology.
Therefore it is only a useful solution for early phases in development, where a USB port might exist on the board, and no disassembly is required to reach the data lines.

\subsubsection{Stop and Radio (Aperiodic Wireless Transmission)}
"Stop and Radio" also saves data continuously in order to be able to stop and transfer the data at once.
At this point, it is also possible to reprogram the car.
However, the aperiodic transmission happens wirelessly via the ESP32's built-in 2.4~GHz wireless capabilities (WiFi 802.11 b/g/n, Bluetooth v4.2, and BLE).
The whole stop/transmission/drive-off process now requires no physical access to the device.

\textbf{Advantages}
\begin{itemize}
    \item Stopping possible anywhere on track (except lane changes)
    \item Virtually unlimited energy available, making the transmission time irrelevant
    \item Updates and development are possible
    \item Low risk of interruption
    \item Low runtime performance impact
\end{itemize}

\textbf{Shortcomings}
\begin{itemize}
    \item Device needs to know when to stop --- physical access might be necessary to, e.g., push a button
    \item Stopping interrupts normal operation --- possibly infeasible in production
    \item WiFi has high flash and RAM memory consumption
    \item Data must be saved whilst driving
\end{itemize}

\textbf{Evaluation}
Altogether, "Stop and Radio" trades off memory and compute power for not having to access the device physically compared to "Save and Print Later".
This makes it especially useful for Over-the-Air (OTA) updates and development, especially at later stages of development.

\subsubsection{Write to Carrera Track (Continuous Wired Transmission)}
\label{subsubsec:Write_to_Carrera_Track}

The Carrera digital slot cars are constantly communicating with the control unit of the track via the tracks' power rails using a digital communication protocol.
This way, multiple cars can be in contact with the same rails and can still be individually controlled.
Slot cars can also send some data back, a feature that is usually used by the car to indicate that it reached a sensor.
There is, however, a possibility that these data packets could be hijacked to send custom data, as much as the protocol allows, back onto the track.
This data could then be read and interpreted by a custom decoder, implementing a continuous, wired transmission.

The advantage would be a constant, reliable data output that is less subject to interference like a radio signal might be.
Using this technique, data to be transferred would not have to be saved for extended periods of time but can be sent with minimal delay.
According to the research by Heß~\cite{SlotBaer}, cars can send up to 13~bits, 8 times per 75~ms cycle.
$$\frac{1}{0.075\textrm{~s}} \cdot 8 \cdot 13\textrm{~Bit} \approx 1,387~\frac{\textrm{Bit}}{\textrm{s}}$$
This results in a theoretical bandwidth of $\approx$ 1,387 Bit/s.

\textbf{Advantages}
\begin{itemize}
    \item Continuous output
    \item Position on track does not matter (except lane changes)
\end{itemize}

\textbf{Shortcomings}
\begin{itemize}
    \item Necessary to adhere to Carrera Digital protocol~\cite{SlotBaer}
    \item Low bandwidth
    \item Difficult to set up --- hardware changes needed on both ends
\end{itemize}
\textbf{Evaluation}
In the end, writing to the Carrera track is difficult to set up, both for developers and potential users and offers only extremely limited bandwidth.
Furthermore, for any mobile embedded devices, similar options are not even available.
For this reason, we did not implement this option.

\subsubsection{WiFi On-the-fly (Continuous Wireless Transmission)}
As wireless transmission is possible from anywhere on a slot car track, it allows for a continuous, direct connection to the slot car.
This way, debug output can be sent at all times and from anywhere on the track.
This is similar to writing to the Carrera track but using a wireless medium.
The challenge with this technique is its high power consumption and performance impact, as analyzed by Pötsch et al.~\cite{DBLP:conf/i2mtc/PotschBS17}.
Hence, energy consumption (here, crossing lane change segments) needs special consideration.

\textbf{Advantages}
\begin{itemize}
    \item Continuous output
    \item Position on track does not matter
    \item OTA updates and development possible
\end{itemize}

\textbf{Shortcomings}
\begin{itemize}
    \item High power consumption
    \begin{itemize}
        \item Problematic on lane changes
        \item Unpredictable spikes in power consumption on send
    \end{itemize}
    \item High memory consumption of WiFi stack
    \item WiFi-stack always active --- can have a runtime impact
    \item OTA updates might be interrupted
    \begin{itemize}
        \item Ejection from track, power loss on lane changes
        \item Measures against data corruption necessary
    \end{itemize}
\end{itemize}

\textbf{Evaluation}
This is the most versatile and convenient option, useful for almost any kind of debugging.
However, high energy consumption, memory usage, and more advanced setup make it the most resource-intensive way of communication.
Special precautions must be taken to prevent the microcontroller from consuming too much power when only limited energy is available.
In the worst case, these power interruptions could result in corrupted data or even a corrupted flash image.

\subsection{Evaluation}
\label{subsec:Evaluation}

Except \textit{Write  to  Carrera  Track}, all the presented debugging techniques have been used in different stages of the software development for the self-driving slot car.
The analysis presented in \Cref{subsec:Debugging_Technique_Application} has been conducted during the aforementioned development process and supports the debugging trade-offs identified in \Cref{sec:Debugging_Techniques_Comparison}.

Though being supported by many microcontrollers, the wired transmission of debug logs can have practical issues:
During the intended mode of operation, the device might need to move.
Therefore, attaching cables is either impossible or distorts the operational characteristics, precluding optimization or production scenarios.
Our particular experiment setup would have allowed the loophole of continuous data transmission via a protocol modulated on the power supply.
However, in addition to a relatively high implementation effort, the protocol constraints limit the communication bandwidth to an extent impractical for development or optimization.

In contrast, wireless transmission of debug messages has proven to offer a broader field of feasible application scenarios.
The available bandwidth enabled the transmission of larger volumes of sensor data and verbose status messages during early application development and optimization activities.
A distorting factor on the devices' normal operating conditions has been the increased power consumption of the radio module.
However, our experiment has shown that careful energy budgeting and adaptive control of power states can overcome this obstacle and avoid interference between the regular device operation and the debugging facilities.

\section{Conclusion and Future Work}

The increased availability of embedded \gls{iot} devices to a wider audience of developers and the increased mobility of such devices calls for more convenient debugging techniques.

In this paper, we report that wireless debugging techniques offer increased applicability and convenience at the price of increased efforts for energy budgeting under constrained circumstances (e.g., energy supply, compute capabilities, waste heat).
To overcome those burdens, readily available hardware and software modules require more adaptive instrumentation.

For instance, future research could focus on techniques to automatically select power states based on a configurable energy budget and control transmission timings accordingly.

\bibliographystyle{IEEEtran}
\bibliography{references}

\begin{thebibliography}{10}
\providecommand{\url}[1]{#1}
\csname url@samestyle\endcsname
\providecommand{\newblock}{\relax}
\providecommand{\bibinfo}[2]{#2}
\providecommand{\BIBentrySTDinterwordspacing}{\spaceskip=0pt\relax}
\providecommand{\BIBentryALTinterwordstretchfactor}{4}
\providecommand{\BIBentryALTinterwordspacing}{\spaceskip=\fontdimen2\font plus
\BIBentryALTinterwordstretchfactor\fontdimen3\font minus
  \fontdimen4\font\relax}
\providecommand{\BIBforeignlanguage}[2]{{%
\expandafter\ifx\csname l@#1\endcsname\relax
\typeout{** WARNING: IEEEtran.bst: No hyphenation pattern has been}%
\typeout{** loaded for the language `#1'. Using the pattern for}%
\typeout{** the default language instead.}%
\else
\language=\csname l@#1\endcsname
\fi
#2}}
\providecommand{\BIBdecl}{\relax}
\BIBdecl

\bibitem{ebert2009embedded}
C.~Ebert and C.~Jones, ``Embedded software: Facts, figures, and future,''
  \emph{Computer}, vol.~42, no.~4, pp. 42--52, 2009.

\bibitem{britton2013reversible}
T.~Britton, L.~Jeng, G.~Carver, P.~Cheak, and T.~Katzenellenbogen, ``Reversible
  debugging software,'' \emph{Judge Bus. School, Univ. Cambridge, Cambridge,
  UK, Tech. Rep}, 2013.

\bibitem{DBLP:journals/sigbed/Mohan08}
\BIBentryALTinterwordspacing
S.~Mohan, ``Worst-case execution time analysis of security policies for deeply
  embedded real-time systems,'' \emph{{SIGBED} Rev.}, vol.~5, no.~1, p.~8,
  2008. [Online]. Available: \url{https://doi.org/10.1145/1366283.1366291}
\BIBentrySTDinterwordspacing

\bibitem{DBLP:conf/isorc/RichterGP15}
\BIBentryALTinterwordspacing
D.~Richter, A.~Grapentin, and A.~Polze, ``Mobility-as-a-service: {A}
  distributed real-time simulation with carrera slot-cars,'' in \emph{{IEEE}
  18th International Symposium on Real-Time Distributed Computing, {ISORC}
  2015, Auckland, New Zealand, 13-17 April, 2015}.\hskip 1em plus 0.5em minus
  0.4em\relax {IEEE} Computer Society, 2015, pp. 276--279. [Online]. Available:
  \url{https://doi.org/10.1109/ISORC.2015.19}
\BIBentrySTDinterwordspacing

\bibitem{macnamee2000emerging}
C.~MacNamee and D.~Heffernan, ``Emerging on-ship debugging techniques for
  real-time embedded systems,'' \emph{Computing \& Control Engineering
  Journal}, vol.~11, no.~6, pp. 295--303, 2000.

\bibitem{DBLP:journals/csur/Stankovic96}
\BIBentryALTinterwordspacing
J.~A. Stankovic, ``Real-time and embedded systems,'' \emph{{ACM} Comput.
  Surv.}, vol.~28, no.~1, pp. 205--208, 1996. [Online]. Available:
  \url{https://doi.org/10.1145/234313.234400}
\BIBentrySTDinterwordspacing

\bibitem{DBLP:conf/ipsn/WhitehouseTTSKJHDC06}
\BIBentryALTinterwordspacing
K.~Whitehouse, G.~Tolle, J.~Taneja, C.~Sharp, S.~Kim, J.~Jeong, J.~Hui,
  P.~Dutta, and D.~E. Culler, ``Marionette: using {RPC} for interactive
  development and debugging of wireless embedded networks,'' in
  \emph{Proceedings of the Fifth International Conference on Information
  Processing in Sensor Networks, {IPSN} 2006, Nashville, Tennessee, USA, April
  19-21, 2006}, J.~A. Stankovic, P.~B. Gibbons, S.~B. Wicker, and J.~A.
  Paradiso, Eds.\hskip 1em plus 0.5em minus 0.4em\relax {ACM}, 2006, pp.
  416--423. [Online]. Available: \url{https://doi.org/10.1145/1127777.1127840}
\BIBentrySTDinterwordspacing

\bibitem{DBLP:conf/emsoft/CuevaBTMS12}
\BIBentryALTinterwordspacing
P.~L. Cueva, A.~Bertaux, A.~Termier, J.~M{\'{e}}haut, and M.~Santana,
  ``Debugging embedded multimedia application traces through periodic pattern
  mining,'' in \emph{Proceedings of the 12th International Conference on
  Embedded Software, {EMSOFT} 2012, part of the Eighth Embedded Systems Week,
  ESWeek 2012, Tampere, Finland, October 7-12, 2012}, A.~Jerraya, L.~P.
  Carloni, F.~Maraninchi, and J.~Regehr, Eds.\hskip 1em plus 0.5em minus
  0.4em\relax {ACM}, 2012, pp. 13--22. [Online]. Available:
  \url{https://doi.org/10.1145/2380356.2380366}
\BIBentrySTDinterwordspacing

\bibitem{MacNamee2000EmergingOD}
C.~MacNamee and D.~Heffernan, ``Emerging on-ship debugging techniques for
  real-time embedded systems,'' \emph{Computing \& Control Engineering
  Journal}, vol.~11, pp. 295--303, 2000.

\bibitem{IBM:journals/Hailpern02}
\BIBentryALTinterwordspacing
B.~Hailpern and P.~Santhanam, ``Software debugging, testing, and
  verification,'' \emph{IBM Syst. J.}, vol.~41, no.~1, p. 4–12, Jan. 2002.
  [Online]. Available: \url{https://doi.org/10.1147/sj.411.0004}
\BIBentrySTDinterwordspacing

\bibitem{murali2021improved}
A.~Murali, H.~Kakarla, and G.~Anitha~Priyadarshini, ``Improved design debugging
  architecture using low power serial communication protocols for signal
  processing applications,'' \emph{International Journal of Speech Technology},
  vol.~24, 06 2021.

\bibitem{DBLP:conf/codes/SamieBH16}
\BIBentryALTinterwordspacing
F.~Samie, L.~Bauer, and J.~Henkel, ``Iot technologies for embedded computing: a
  survey,'' in \emph{Proceedings of the Eleventh {IEEE/ACM/IFIP} International
  Conference on Hardware/Software Codesign and System Synthesis, {CODES} 2016,
  Pittsburgh, Pennsylvania, USA, October 1-7, 2016}.\hskip 1em plus 0.5em minus
  0.4em\relax {ACM}, 2016, pp. 8:1--8:10. [Online]. Available:
  \url{https://doi.org/10.1145/2968456.2974004}
\BIBentrySTDinterwordspacing

\bibitem{yue2013marine}
M.~Yue and Y.~Sun, ``Marine data collection based on embedded system with wired
  and wireless transmission,'' Master's thesis, Universitetet i
  Agder/University of Agder, 2013.

\bibitem{DBLP:conf/infocom/FriedmanKK11}
\BIBentryALTinterwordspacing
R.~Friedman, A.~Kogan, and Y.~Krivolapov, ``On power and throughput tradeoffs
  of wifi and bluetooth in smartphones,'' in \emph{{INFOCOM} 2011. 30th {IEEE}
  International Conference on Computer Communications, Joint Conference of the
  {IEEE} Computer and Communications Societies, 10-15 April 2011, Shanghai,
  China}.\hskip 1em plus 0.5em minus 0.4em\relax {IEEE}, 2011, pp. 900--908.
  [Online]. Available: \url{https://doi.org/10.1109/INFCOM.2011.5935315}
\BIBentrySTDinterwordspacing

\bibitem{DBLP:conf/i2mtc/PotschBS17}
\BIBentryALTinterwordspacing
A.~Potsch, A.~Berger, and A.~Springer, ``Efficient analysis of power
  consumption behaviour of embedded wireless iot systems,'' in \emph{{IEEE}
  International Instrumentation and Measurement Technology Conference, {I2MTC}
  2017, Torino, Italy, May 22-25, 2017}.\hskip 1em plus 0.5em minus 0.4em\relax
  {IEEE}, 2017, pp. 1--6. [Online]. Available:
  \url{https://doi.org/10.1109/I2MTC.2017.7969658}
\BIBentrySTDinterwordspacing

\bibitem{DBLP:conf/hpec/SkowyraBB13}
\BIBentryALTinterwordspacing
R.~Skowyra, S.~Bahargam, and A.~Bestavros, ``Software-defined {IDS} for
  securing embedded mobile devices,'' in \emph{{IEEE} High Performance Extreme
  Computing Conference, {HPEC} 2013, Waltham, MA, USA, September 10-12,
  2013}.\hskip 1em plus 0.5em minus 0.4em\relax {IEEE}, 2013, pp. 1--7.
  [Online]. Available: \url{https://doi.org/10.1109/HPEC.2013.6670325}
\BIBentrySTDinterwordspacing

\bibitem{DBLP:conf/aspdac/HwangKJC08}
\BIBentryALTinterwordspacing
Y.~Hwang, S.~Ku, C.~Jung, and K.~Chung, ``Predictive power aware management for
  embedded mobile devices,'' in \emph{Proceedings of the 13th Asia South
  Pacific Design Automation Conference, {ASP-DAC} 2008, Seoul, Korea, January
  21-24, 2008}, C.~Kyung, K.~Choi, and S.~Ha, Eds.\hskip 1em plus 0.5em minus
  0.4em\relax {IEEE}, 2008, pp. 36--41. [Online]. Available:
  \url{https://doi.org/10.1109/ASPDAC.2008.4483976}
\BIBentrySTDinterwordspacing

\bibitem{5751382}
A.~Ukil, J.~Sen, and S.~Koilakonda, ``Embedded security for internet of
  things,'' in \emph{2011 2nd National Conference on Emerging Trends and
  Applications in Computer Science}, 2011, pp. 1--6.

\bibitem{10.1145/1435458.1435462}
\BIBentryALTinterwordspacing
J.~Brakensiek, A.~Dr\"{o}ge, M.~Botteck, H.~H\"{a}rtig, and A.~Lackorzynski,
  ``Virtualization as an enabler for security in mobile devices,'' in
  \emph{Proceedings of the 1st Workshop on Isolation and Integration in
  Embedded Systems}, ser. IIES '08.\hskip 1em plus 0.5em minus 0.4em\relax New
  York, NY, USA: Association for Computing Machinery, 2008, p. 17–22.
  [Online]. Available: \url{https://doi.org/10.1145/1435458.1435462}
\BIBentrySTDinterwordspacing

\bibitem{DBLP:conf/uist/DrewNMMMH16}
\BIBentryALTinterwordspacing
D.~Drew, J.~L. Newcomb, W.~McGrath, F.~Maksimovic, D.~Mellis, and B.~Hartmann,
  ``The toastboard: Ubiquitous instrumentation and automated checking of
  breadboarded circuits,'' in \emph{Proceedings of the 29th Annual Symposium on
  User Interface Software and Technology, {UIST} 2016, Tokyo, Japan, October
  16-19, 2016}, J.~Rekimoto, T.~Igarashi, J.~O. Wobbrock, and D.~Avrahami,
  Eds.\hskip 1em plus 0.5em minus 0.4em\relax {ACM}, 2016, pp. 677--686.
  [Online]. Available: \url{https://doi.org/10.1145/2984511.2984566}
\BIBentrySTDinterwordspacing

\bibitem{DBLP:conf/chi/BoothSBJ16}
\BIBentryALTinterwordspacing
T.~Booth, S.~Stumpf, J.~Bird, and S.~Jones, ``Crossed wires: Investigating the
  problems of end-user developers in a physical computing task,'' in
  \emph{Proceedings of the 2016 {CHI} Conference on Human Factors in Computing
  Systems, San Jose, CA, USA, May 7-12, 2016}, J.~Kaye, A.~Druin, C.~Lampe,
  D.~Morris, and J.~P. Hourcade, Eds.\hskip 1em plus 0.5em minus 0.4em\relax
  {ACM}, 2016, pp. 3485--3497. [Online]. Available:
  \url{https://doi.org/10.1145/2858036.2858533}
\BIBentrySTDinterwordspacing

\bibitem{DBLP:conf/uist/McGrathWKHDH18}
\BIBentryALTinterwordspacing
W.~McGrath, J.~Warner, M.~Karchemsky, A.~Head, D.~Drew, and B.~Hartmann,
  ``Wifr{\"{o}}st: Bridging the information gap for debugging of networked
  embedded systems,'' in \emph{The 31st Annual {ACM} Symposium on User
  Interface Software and Technology, {UIST} 2018, Berlin, Germany, October
  14-17, 2018}, P.~Baudisch, A.~Schmidt, and A.~Wilson, Eds.\hskip 1em plus
  0.5em minus 0.4em\relax {ACM}, 2018, pp. 447--455. [Online]. Available:
  \url{https://doi.org/10.1145/3242587.3242668}
\BIBentrySTDinterwordspacing

\bibitem{DBLP:conf/osdi/FonsecaDLS08}
\BIBentryALTinterwordspacing
R.~Fonseca, P.~Dutta, P.~Levis, and I.~Stoica, ``Quanto: Tracking energy in
  networked embedded systems,'' in \emph{8th {USENIX} Symposium on Operating
  Systems Design and Implementation, {OSDI} 2008, December 8-10, 2008, San
  Diego, California, USA, Proceedings}, R.~Draves and R.~van Renesse,
  Eds.\hskip 1em plus 0.5em minus 0.4em\relax {USENIX} Association, 2008, pp.
  323--338. [Online]. Available:
  \url{http://www.usenix.org/events/osdi08/tech/full_papers/fonseca/fonseca.pdf}
\BIBentrySTDinterwordspacing

\bibitem{DBLP:journals/isj/BaskervilleP04}
\BIBentryALTinterwordspacing
R.~L. Baskerville and J.~Pries{-}Heje, ``Short cycle time systems
  development,'' \emph{Inf. Syst. J.}, vol.~14, no.~3, pp. 237--264, 2004.
  [Online]. Available: \url{https://doi.org/10.1111/j.1365-2575.2004.00171.x}
\BIBentrySTDinterwordspacing

\bibitem{DBLP:journals/ubiquity/Hyde06}
\BIBentryALTinterwordspacing
R.~Hyde, ``The fallacy of premature optimization,'' \emph{Ubiquity}, vol. 2006,
  no. June, p. 4:2, 2006. [Online]. Available:
  \url{https://doi.org/10.1145/1147991.1147993}
\BIBentrySTDinterwordspacing

\bibitem{DBLP:journals/spe/DoddR92}
\BIBentryALTinterwordspacing
P.~S. Dodd and C.~V. Ravishankar, ``Monitoring and debugging distributed
  real-time programs,'' \emph{Softw. Pract. Exp.}, vol.~22, no.~10, pp.
  863--877, 1992. [Online]. Available:
  \url{https://doi.org/10.1002/spe.4380221005}
\BIBentrySTDinterwordspacing

\bibitem{DBLP:journals/cacm/Spinellis18}
\BIBentryALTinterwordspacing
D.~Spinellis, ``Modern debugging: the art of finding a needle in a haystack,''
  \emph{Commun. {ACM}}, vol.~61, no.~11, pp. 124--134, 2018. [Online].
  Available: \url{https://doi.org/10.1145/3186278}
\BIBentrySTDinterwordspacing

\bibitem{DBLP:conf/www/QianWGHGMSS12}
\BIBentryALTinterwordspacing
F.~Qian, Z.~Wang, Y.~Gao, J.~Huang, A.~Gerber, Z.~M. Mao, S.~Sen, and
  O.~Spatscheck, ``Periodic transfers in mobile applications: network-wide
  origin, impact, and optimization,'' in \emph{Proceedings of the 21st World
  Wide Web Conference 2012, {WWW} 2012, Lyon, France, April 16-20, 2012},
  A.~Mille, F.~Gandon, J.~Misselis, M.~Rabinovich, and S.~Staab, Eds.\hskip 1em
  plus 0.5em minus 0.4em\relax {ACM}, 2012, pp. 51--60. [Online]. Available:
  \url{https://doi.org/10.1145/2187836.2187844}
\BIBentrySTDinterwordspacing

\bibitem{DBLP:conf/etfa/SenoVT11}
\BIBentryALTinterwordspacing
L.~Seno, S.~Vitturi, and F.~Tramarin, ``Experimental evaluation of the service
  time for industrial hybrid (wired/wireless) networks under non-ideal
  environmental conditions,'' in \emph{{IEEE} 16th Conference on Emerging
  Technologies {\&} Factory Automation, {ETFA} 2011, Toulouse, France,
  September 5-9, 2011}, Z.~Mammeri, Ed.\hskip 1em plus 0.5em minus 0.4em\relax
  {IEEE}, 2011, pp. 1--8. [Online]. Available:
  \url{https://doi.org/10.1109/ETFA.2011.6059005}
\BIBentrySTDinterwordspacing

\bibitem{DBLP:journals/tvlsi/ZhaiPNHORMHASB09}
\BIBentryALTinterwordspacing
B.~Zhai, S.~Pant, L.~Nazhandali, S.~Hanson, J.~Olson, A.~Reeves, M.~Minuth,
  R.~Helfand, T.~M. Austin, D.~Sylvester, and D.~T. Blaauw, ``Energy-efficient
  subthreshold processor design,'' \emph{{IEEE} Trans. Very Large Scale Integr.
  Syst.}, vol.~17, no.~8, pp. 1127--1137, 2009. [Online]. Available:
  \url{https://doi.org/10.1109/TVLSI.2008.2007564}
\BIBentrySTDinterwordspacing

\bibitem{ESP32DataSheet}
{Espressif Systems}, ``Esp32 series datasheet,''
  \url{https://www.espressif.com/sites/default/files/documentation/esp32_datasheet_en.pdf},
  2021, accessed: 2021-08-09.

\bibitem{SlotBaer}
S.~Heß, ``Slotbaer / carrera digital 124/132 / cu daten-protokoll (update
  2019),''
  \url{http://slotbaer.de/carrera-digital-124-132/9-cu-daten-protokoll.html},
  2019, accessed: 2021-05-18.

\end{thebibliography}

\end{document}